\documentclass[usenatbib,useAMS]{mn2e}

\usepackage{rotating}
\usepackage{graphicx}
\usepackage{amsmath}
\usepackage{amssymb}
\usepackage{bbold}
\usepackage{multirow}
\usepackage{url}
 
\usepackage{xcolor}

\usepackage{float}

\newcommand{\kms}{\mbox{km\,s$^{-1}$}}
\newcommand{\kuns}{\mbox{$h$\,Mpc$^{-1}$}}
\newcommand{\Muns}{\mbox{$h^{-1}\,{\rm Mpc}$}}
\newcommand{\degree}{\mbox{$^{\circ}$}}

\newcommand{\mbi}[1]{\mbox{\boldmath$#1$}}

\newcommand{\lsim}{\mbox{${\,\hbox{\hbox{$ < $}\kern -0.8em \lower 1.0ex\hbox{$\sim$}}\,}$}}
\newcommand{\gsim}{\mbox{${\,\hbox{\hbox{$ > $}\kern -0.8em \lower 1.0ex\hbox{$\sim$}}\,}$}}

\def\etal{{\it et al.\ }}

\def\beqn{\vspace{2mm}
\begin{eqnarray}} 
\def\eeqn{\vspace{2mm} 
\end{eqnarray}}

 \def\la{\langle}

\newcommand{\be}{\begin{equation}}
\newcommand{\ee}{\end{equation}}
\newcommand{\ba}{\begin{eqnarray}}
\newcommand{\ea}{\end{eqnarray}}
\newcommand{\brr}{\begin{array}}
 
\newcommand{\err}{\end{array}}
\newcommand{\bc}{\begin{center}}
\newcommand{\ec}{\end{center}}

\begin{document}

\title[Cosmic Structure and Dynamics]{Cosmic Structure and Dynamics of the Local Universe}

\author[Kitaura \etal]{Francisco-Shu Kitaura$^{1}$\thanks{E-mail: kitaura@aip.de, Karl-Schwarzschild fellow}, Pirin Erdo\u{g}du$^2$,  Sebasti{\'a}n E.~Nuza$^1$, Arman Khalatyan$^1$, \and  Raul E.~Angulo$^3$, Yehuda Hoffman$^4$ and Stefan Gottl{\"o}ber$^1$\\$^{1}$  Leibniz-Institut f\"ur Astrophysik (AIP), An der Sternwarte 16, D-14482 Potsdam, Germany \\$^{2}$ Department of Physics and Astronomy, University College London, London WC1E 6BT, United Kingdom  \\$^{3}$  Max-Planck Institut f\"ur Astrophysik (MPA), Karl-Schwarzschildstr.~1, D-85748 Garching, Germany \\$^{4}$  Racah Institute of Physics, Hebrew University, Jerusalem, Israel}

\maketitle

\begin{abstract}
We present a cosmography analysis of the Local Universe based on the recently released Two-Micron All-Sky Redshift Survey (2MRS). 
Our method is based on a Bayesian Networks Machine Learning algorithm (the \textsc{Kigen}-code) which self-consistently samples the initial density fluctuations compatible with the observed galaxy distribution and a structure formation model given by second order Lagrangian perturbation theory (2LPT). From the initial conditions we obtain an ensemble of reconstructed density and peculiar velocity fields which  characterize the local cosmic structure with high accuracy unveiling nonlinear structures like filaments and voids in detail.
 Coherent redshift space distortions are consistently corrected within 2LPT.
 From the ensemble of cross-correlations between the  reconstructions and the galaxy field and the variance of the recovered density fields we find that our method is  extremely accurate up to $k\sim$ 1 \kuns and still yields reliable results { down to scales of about 3-4 {\Muns}}. 
The motion of the local group we obtain within $\sim$ 80 {\Muns}  ($v_{\rm LG}=522\pm86$ \kms, $l_{\rm LG}=291\degree\pm16\degree, b_{\rm LG}=34\degree\pm8\degree$) is in  good agreement with measurements derived from the CMB and from direct observations of peculiar motions and is consistent with the predictions of $\Lambda$CDM.

\end{abstract}

\begin{keywords}
(cosmology:) large-scale structure of Universe -- galaxies: clusters: general --
 catalogues -- galaxies: statistics
\end{keywords}

\section{Introduction}

  The Local Universe (LU) harbours the link between the early Universe and our present day cosmic environment. A profound analysis of its cosmic structure is thus essential to gain insight into the processes which lead to structure formation in our surroundings  and ultimately to our own Galaxy.  This is one of the main goals performing constrained simulations of the LU \citep[see e.~g.][]{1998ApJ...492..439B,clues2,klypin}. 
To carry out this kind of studies one needs first to recover the initial fluctuations which gave rise to the galaxy distribution we observe in our neighbourhood.
However, such a task implies various complications since in general a galaxy redshift survey provides a discrete sample of biased matter tracers which are degraded by observational effects like a radial selection function, due to a magnitude limit cut, and redshift-space distortions caused by peculiar motions with respect to the Hubble flow \citep[see discussion in][and references therein]{2011MNRAS.416.2494P}. For very nearby structures one may use measurements of radial velocities to partially avoid these problems \citep[see e.~g.][]{zaroubi_vel}.  However, one still needs  to trace the observations back in time unfolding gravity which couples matter on different scales in a nonlinear and nonlocal way. While linear approaches are useful to reconstruct the baryon acoustic oscillations signal on very large scales \citep[][]{2007ApJ...664..675E}, they fail on small scales when the nonlinear regime becomes relevant.
One may solve the boundary problem of finding the initial Lagrangian positions of galaxies by minimizing an action as suggested in different works \citep[see e.~g.][]{Peebles89,NB00,BEN02,2003MNRAS.346..501B}. 
The determination of the initial conditions (and hence of the displacement field) automatically yields estimates on the peculiar velocity field  \citep[see also][in addition to the previously cited works]{2005ApJ...635L.113M,LMCTBS08}.
The mismatch in the measurement of the Local Group (LG) velocity from the cosmic microwave background (CMB) \citep[][]{wmap5} and from the matter distribution in the LU \citep[see][and references therein]{bilicki}
together with the recent claims on large-scale flows which could challenge the standard cosmological model \citep[{ or give some hints on the initial perturbations of the Universe, see e.~g.}][]{2008ApJ...686L..49K,2009MNRAS.392..743W} stresses the need for more accurate studies of the local dynamics.
 It should be noted that the number of solutions compatible with the observations is degenerate due to shell crossings and redshift-space distortions  \citep[see e.~g.][]{yahil}.
 Moreover, the approaches mentioned above do not provide yet the initial fluctuation field. At this point another assumption needs to be made on the statistics of the initial fluctuations. 
From these arguments it should be clear that the estimation of the initial conditions corresponding to a galaxy distribution in redshift-space has a stochastic nature which should be treated in a statistical way to accurately model the propagation of uncertainties. 
 For this reason we suggest to extend the Bayesian works based on linear Gaussian fields \citep[{ see the pioneering works during the 90s:}][]{hoffman,zaroubi,fisher,rien,zaroubi_vel,inga}  and apply a Bayesian Networks Machine Learning approach including a nonlinear and nonlocal model for structure formation providing an ensemble of reconstructed initial and final density and peculiar velocity fields \citep[see][]{kitaura_kigen}. 
This ensemble of solutions enables us to estimate in a realistic way the uncertainties in the reconstruction and cross-check the accuracy of the method with the observations, improving previous work concerning single ``optimal'' solutions   limited to  linear Eulerian or Lagrangian perturbation theory.

\begin{figure}
\begin{tabular}{cc}
\hspace{-0.4cm}
\includegraphics[width=4.5cm]{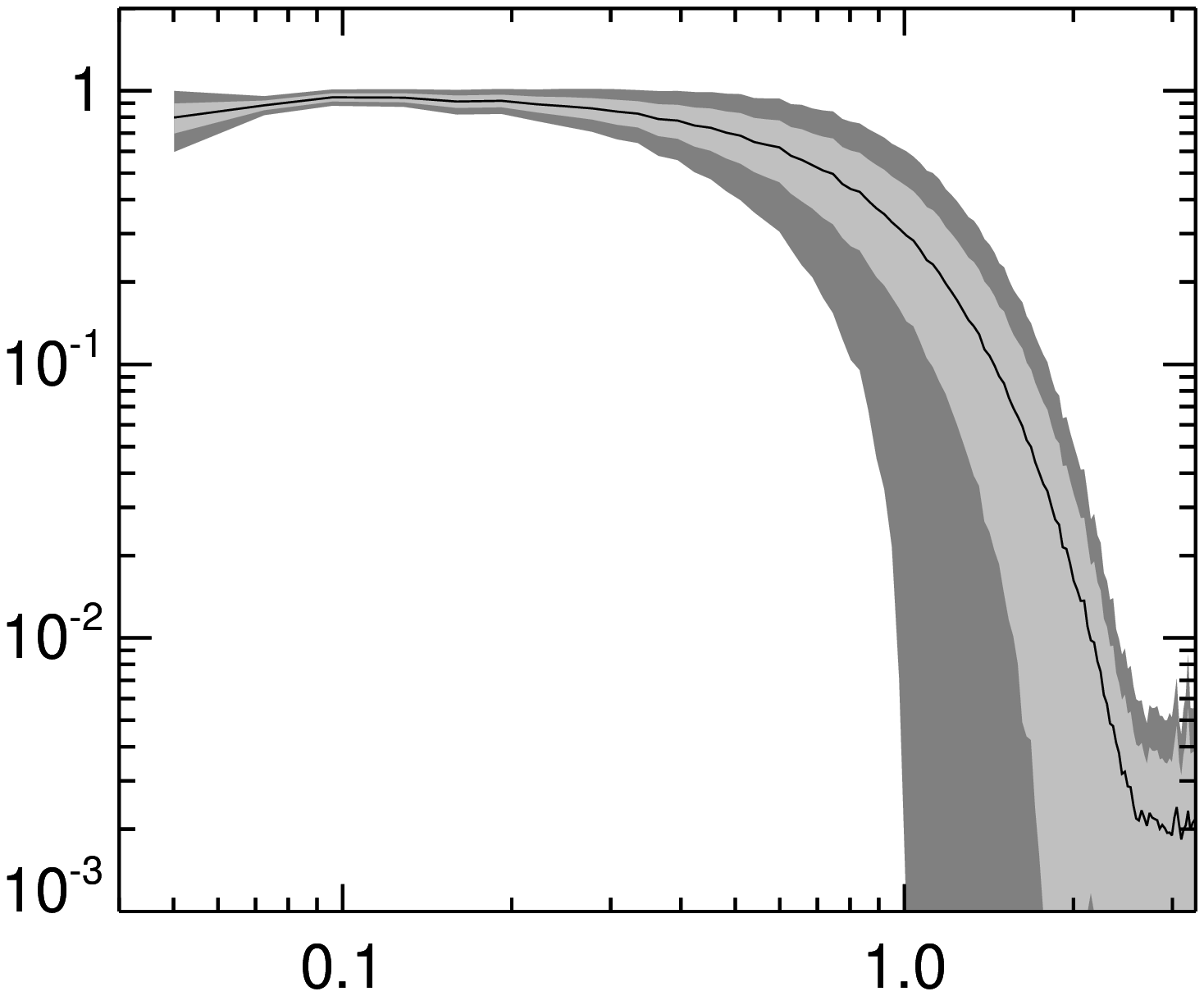}
\put(-137,60){\rotatebox[]{90}{\rm $XP(k)$}}
\put(-85,-5){\rm $k$ [$h$ Mpc$^{-1}$]}
\hspace{-0.5cm}
\includegraphics[width=4.5cm]{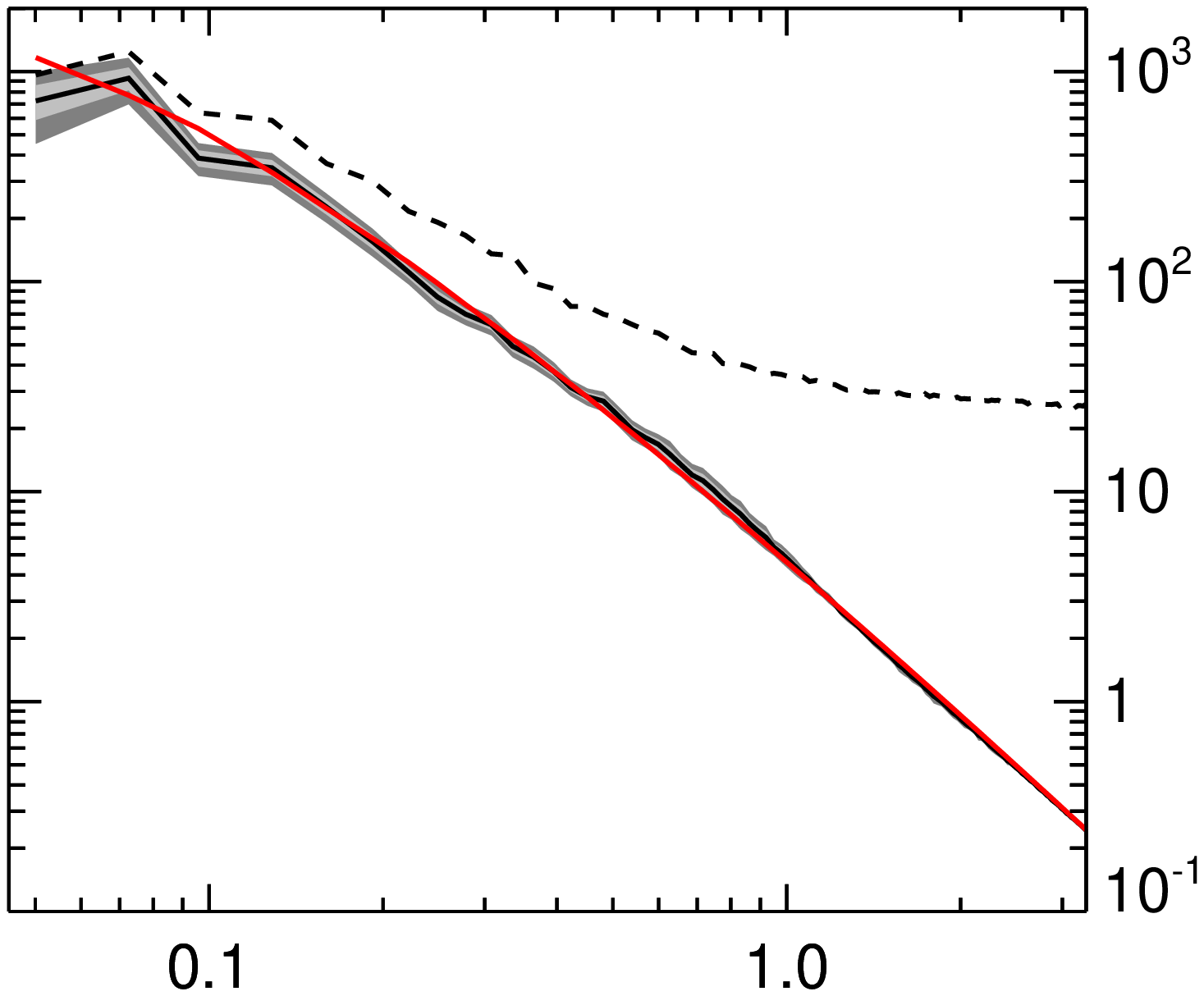}
\put(-4,60){\rotatebox[]{-90}{\rm $P(k)$}}
\put(-93,-5){\rm $k$ [$h$ Mpc$^{-1}$]}
\end{tabular}
\vspace{-0.3cm}
\caption{\label{fig:corr}
{\it On the left}: Normalised cross-power spectra $XP(k)\equiv\langle|\hat{\delta}^{\rm rec}(\mbi k)\overline{\hat{\delta}^{\rm gal}(\mbi k)}|\rangle/(\sqrt{P^{\rm rec}(k)}\sqrt{P^{\rm gal}(k)})$  between the galaxy overdensity and the reconstructed density field  with 1 and 2 sigma contours (light and dark shaded regions, respectively). {\it On the right}: power spectrum of the galaxy field $P^{\rm gal}(k)$ (dashed curve), linear $\Lambda$CDM power spectrum (red curve), mean  of the 100 reconstructed linear power-spectra of the initial field (black curve) with 1 and 2 sigma contours (light and dark shaded regions, respectively).} 
\end{figure}

\section{Data, method and results}

\begin{figure*}
\begin{tabular}{ccc}
\put(-10,95){\rotatebox[]{90}{\rm SGY [$h^{-1}$ Mpc]}}
\includegraphics[width=16.5cm]{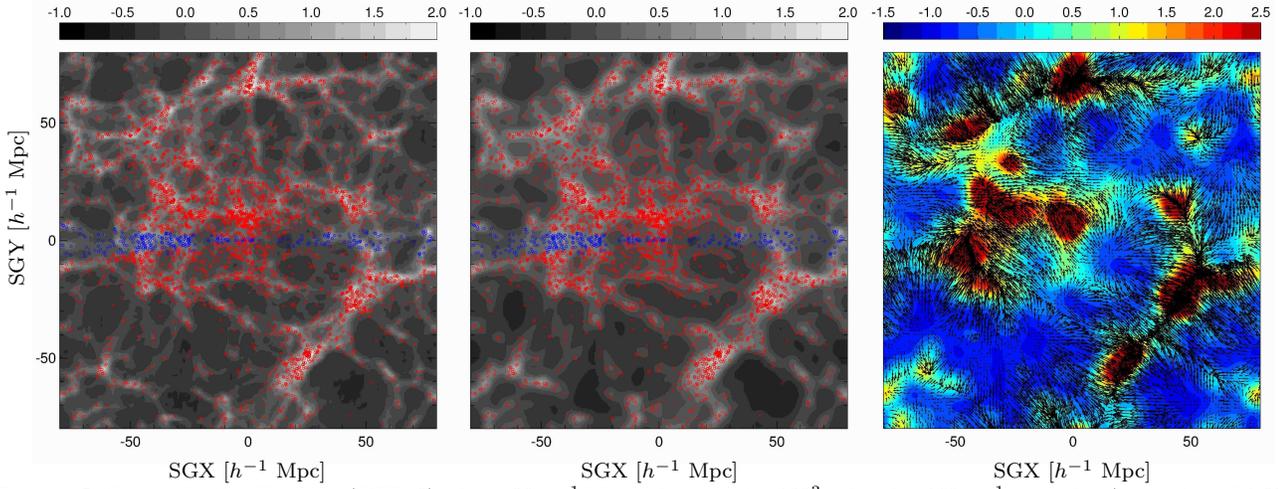}
\put(-418,-5){\rm SGX [$h^{-1}$ Mpc]}
\put(-262,-5){\rm SGX [$h^{-1}$ Mpc]}
\put(-106,-5){\rm SGX [$h^{-1}$ Mpc]}
\end{tabular}
\vspace{-0.2cm}
\caption{\label{fig:SG}
Supergalactic XY-plane (SGZ=0)  with $\sim$20  $h^{-1}$ Mpc thickness of a 128$^3$ grid with 160  $h^{-1}$ Mpc side (resolution of 1.25 $h^{-1}$ Mpc): {\it left panel}:  logarithm of the reconstructed density field { $\ln(2+\delta^{\rm rec})$} (one sample belonging to the highly correlated subsample of 21 reconstructions which have a cross-correlation with the galaxy overdensity better than 1 sigma at scales $\ge$ 3.5 {\Muns}) with overplotted observed galaxies in red and augmented ones in blue; {\it middle panel}: logarithm of the mean density field of all 100 samples { $\ln(2+\langle\delta^{\rm rec}\rangle)$}, {\it right panel}: $v_x-v_y$ velocity field with the underlying galaxy overdensity field after 3.5   $h^{-1}$ Mpc Gaussian smoothing. The length of the arrows is proportional to the average speed at that location.
 } 
\end{figure*}

\begin{figure*}
\begin{tabular}{ccc}
\put(-10,95){\rotatebox[]{90}{\rm SGY [$h^{-1}$ Mpc]}}
\hspace{-0.120cm}
\includegraphics[width=16.75cm]{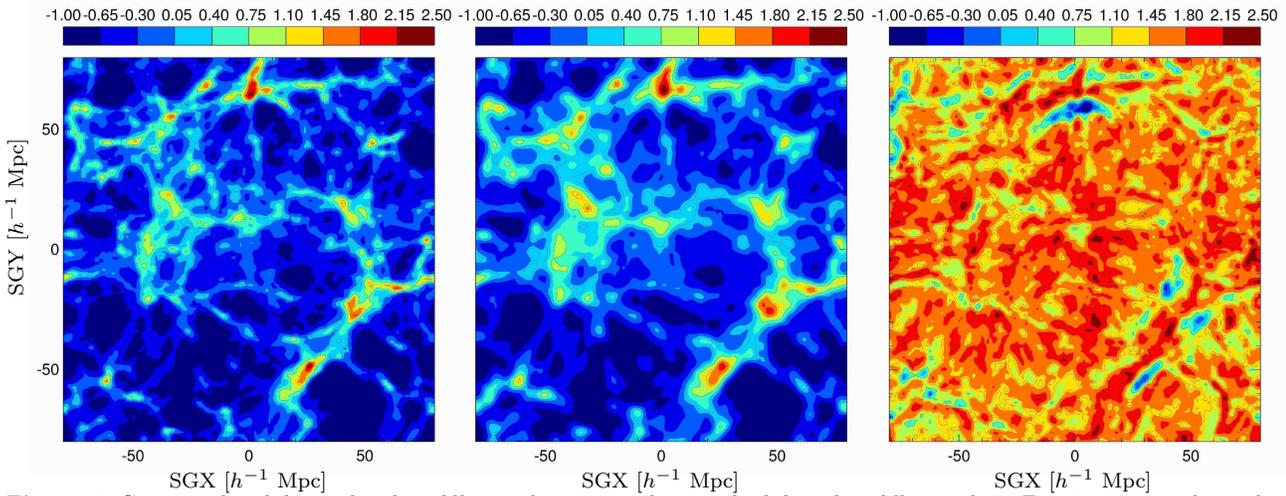}
\put(-424,-5){\rm SGX [$h^{-1}$ Mpc]}
\put(-268,-5){\rm SGX [$h^{-1}$ Mpc]}
\put(-112,-5){\rm SGX [$h^{-1}$ Mpc]}
\end{tabular}
\vspace{-0.2cm}
\caption{\label{fig:CSG} Contour plot: {\it left panel} and {\it middle panel}:  corresponding to the left and middle panels in Fig.~\ref{fig:SG}, respectively; {\it right panel}: logarithm of the signal-to-noise ratio { of the mean density field  (S/N$\equiv\langle\delta^{\rm rec}\rangle/\sigma(\delta^{\rm rec}))$} in each cell from all samples in the same slice as the other panels.
 } 
\end{figure*}

\begin{figure*}
\begin{tabular}{cc}
\rotatebox[]{90}{\includegraphics[width=3.5cm]{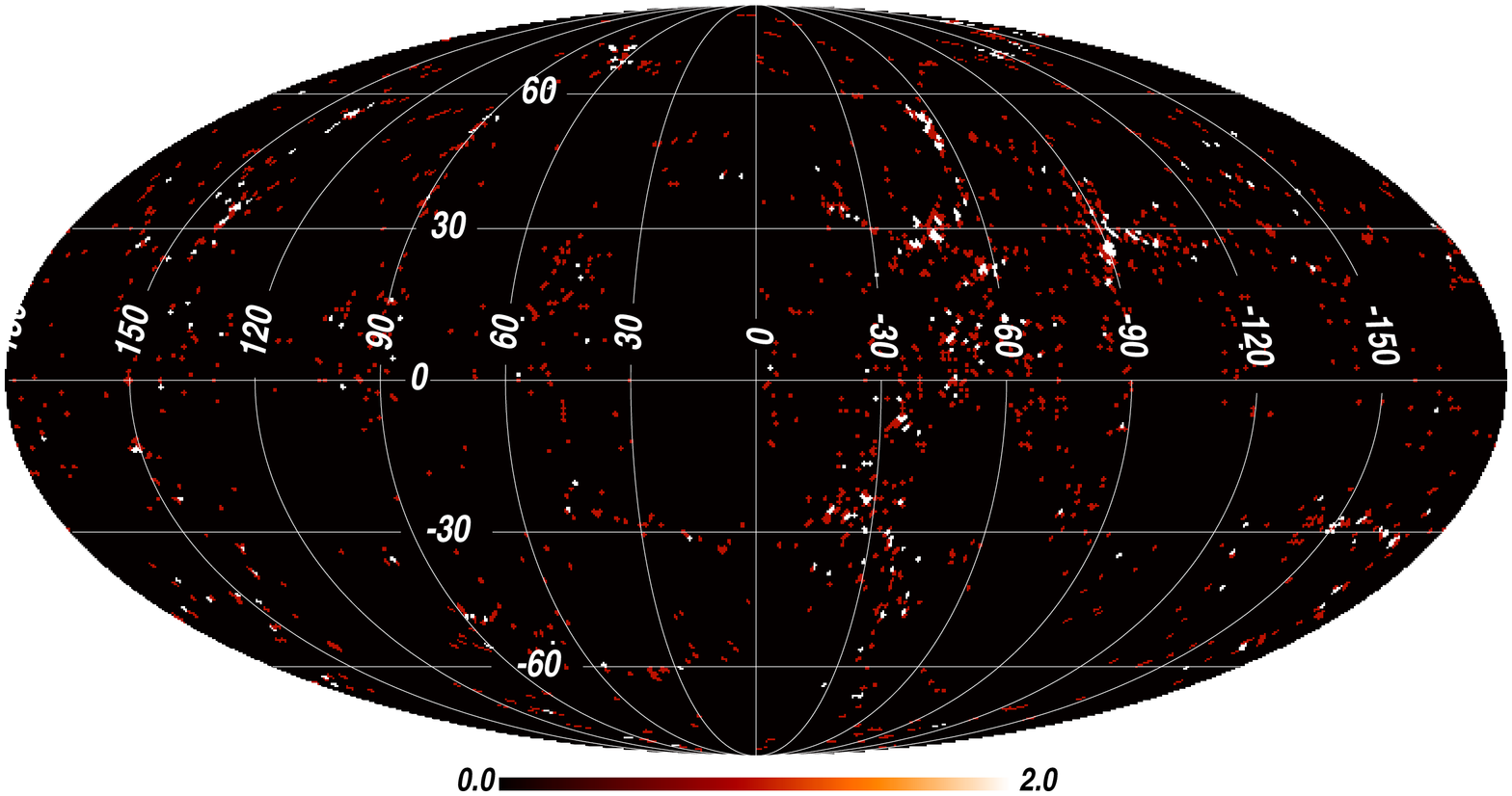}}
\put(60,121){\rm 30-40 $h^{-1}$ Mpc}
\rotatebox[]{90}{\includegraphics[width=3.5cm]{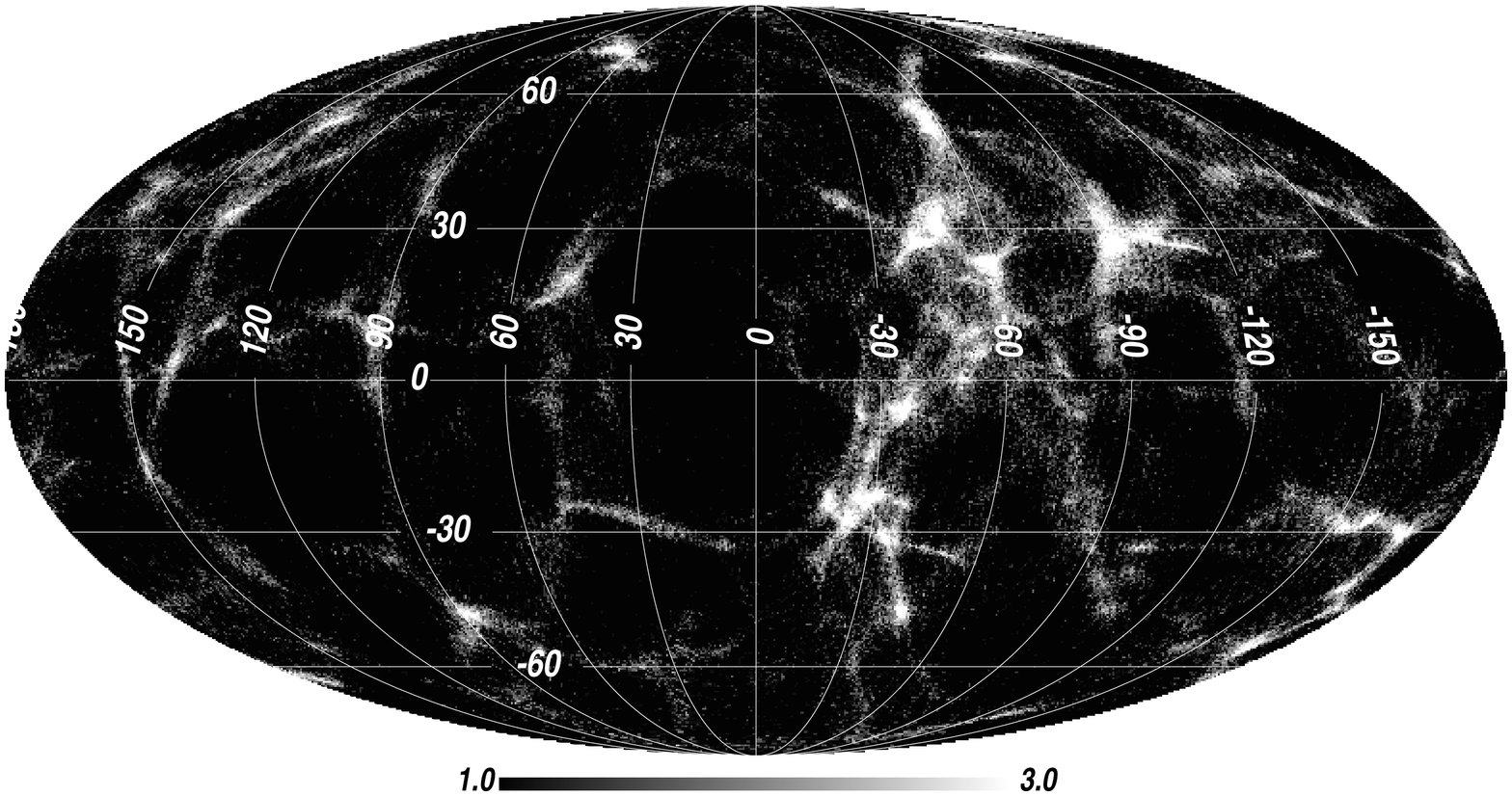}}
\rotatebox[]{90}{\includegraphics[width=3.5cm]{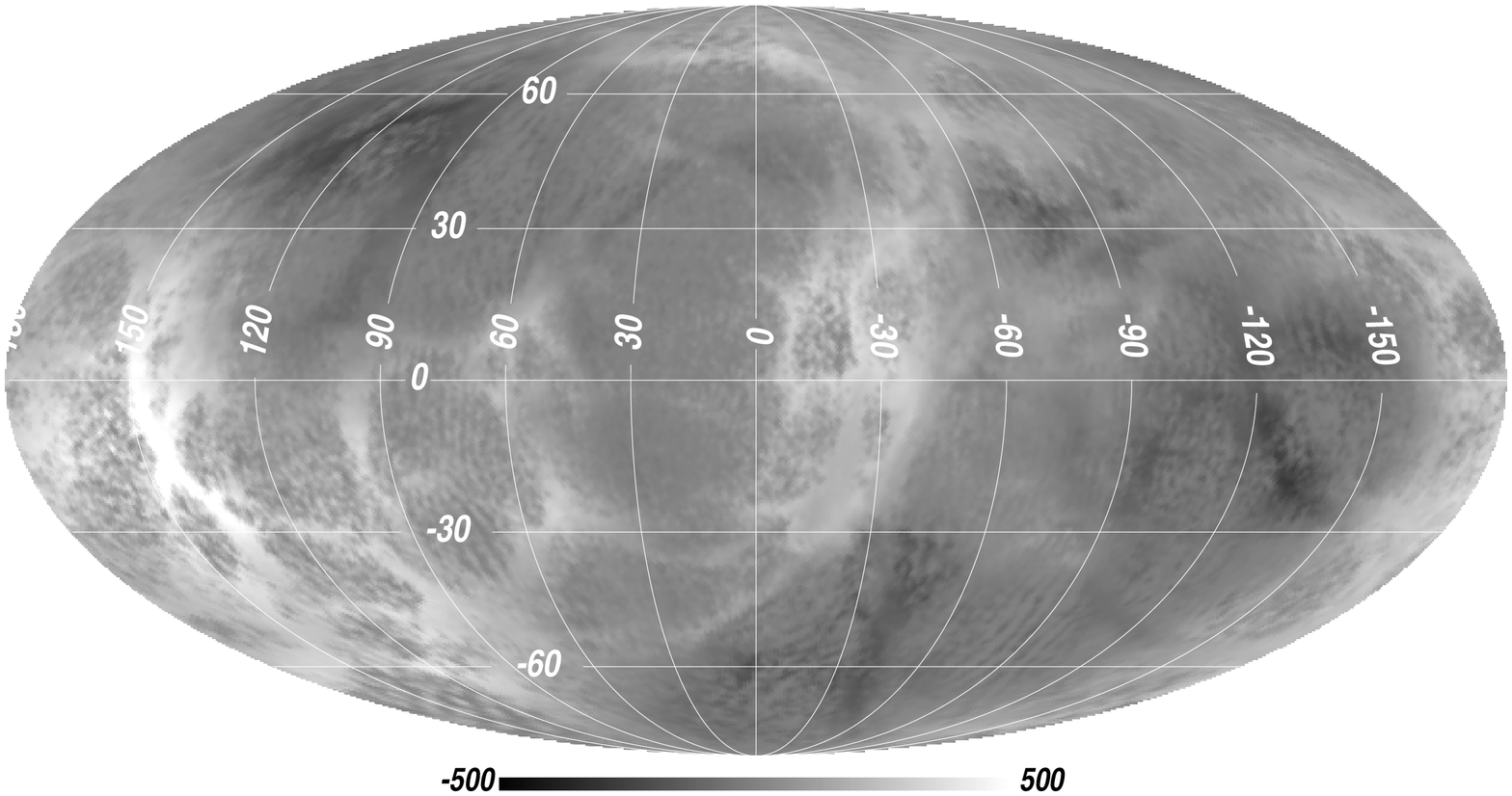}}
\vspace{-1.3cm}
\\
\rotatebox[]{90}{\includegraphics[width=3.5cm]{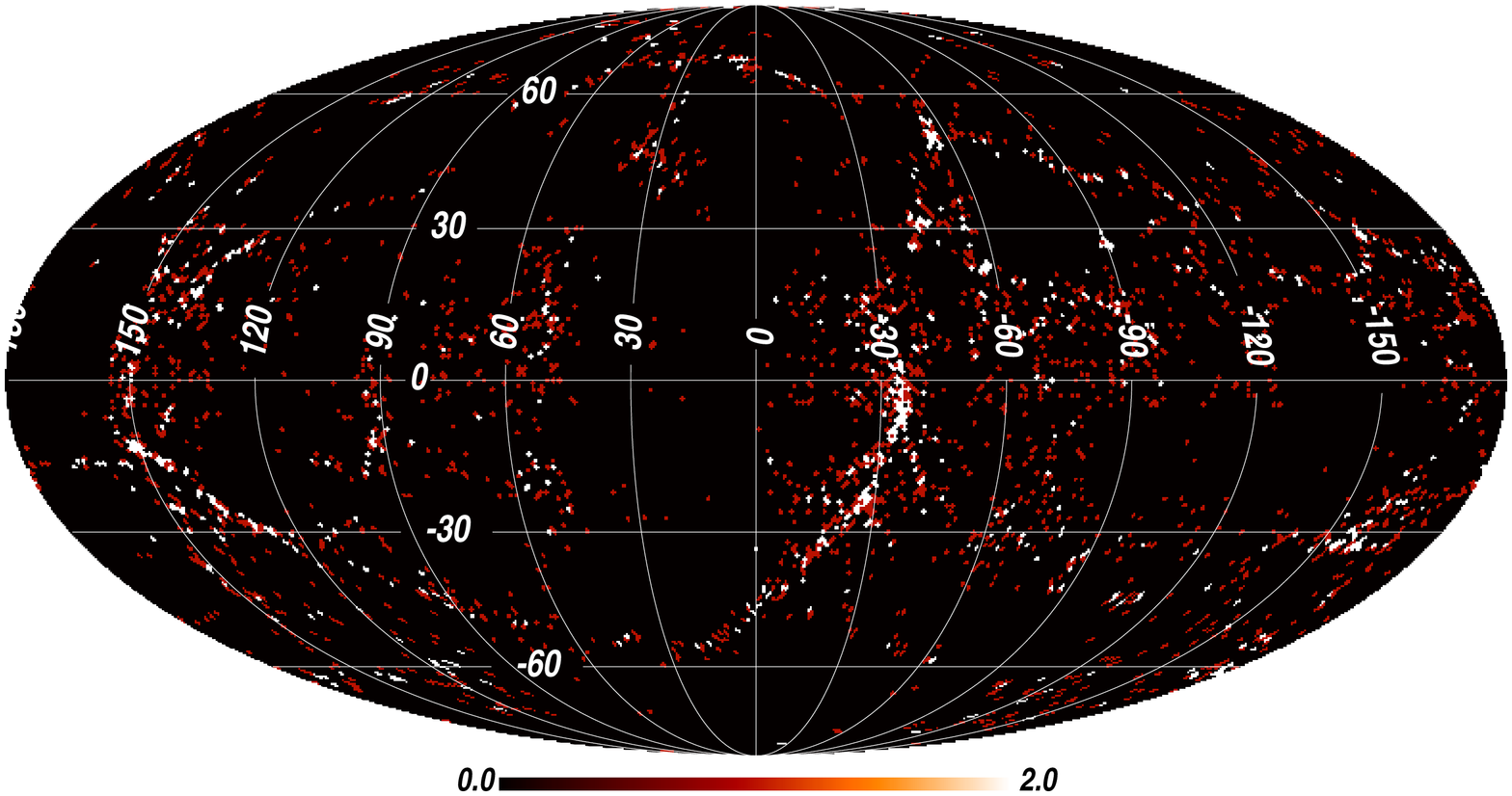}}
\put(60,121){\rm 40-50 $h^{-1}$ Mpc}
\rotatebox[]{90}{\includegraphics[width=3.5cm]{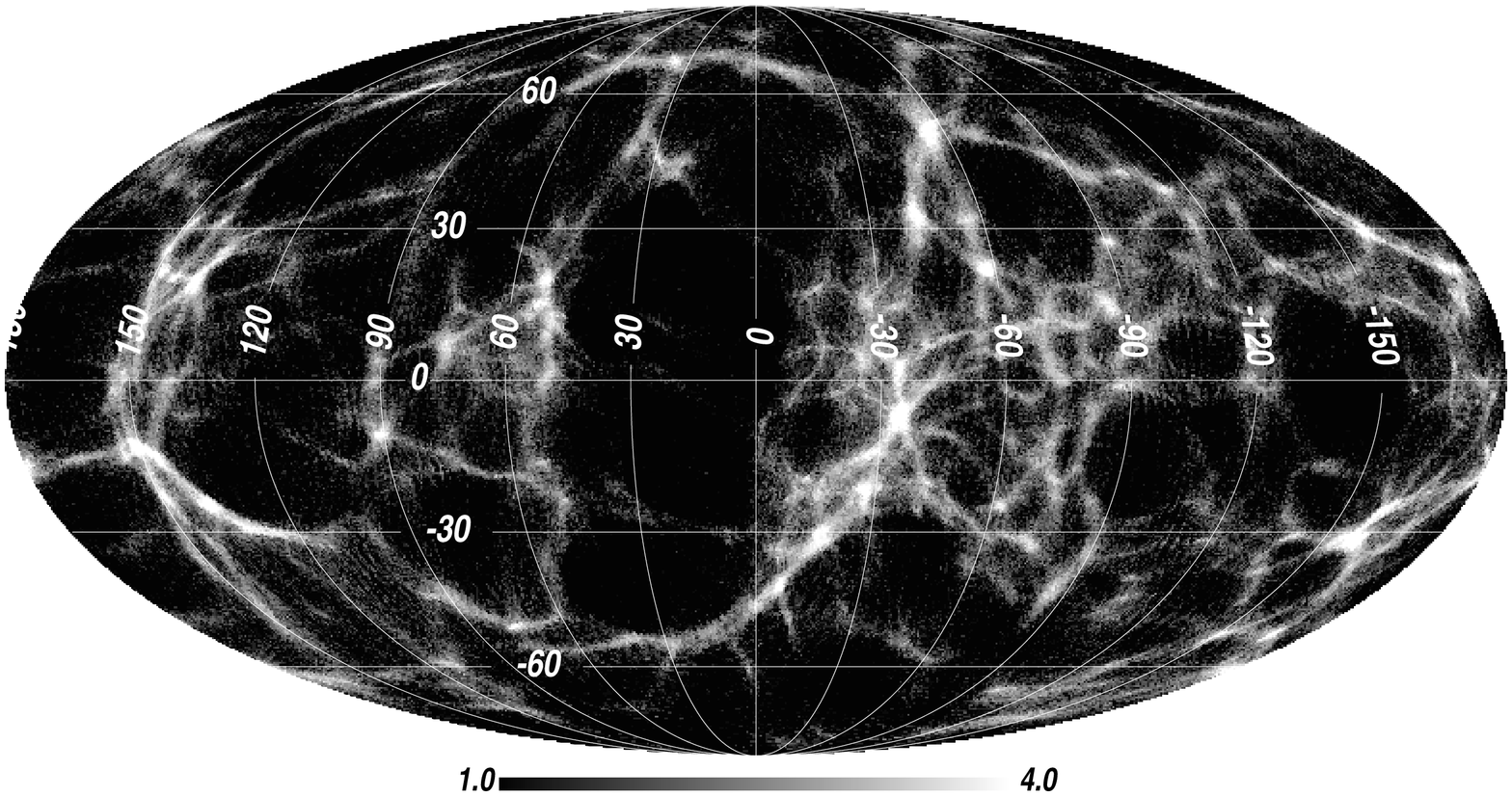}}
\rotatebox[]{90}{\includegraphics[width=3.5cm]{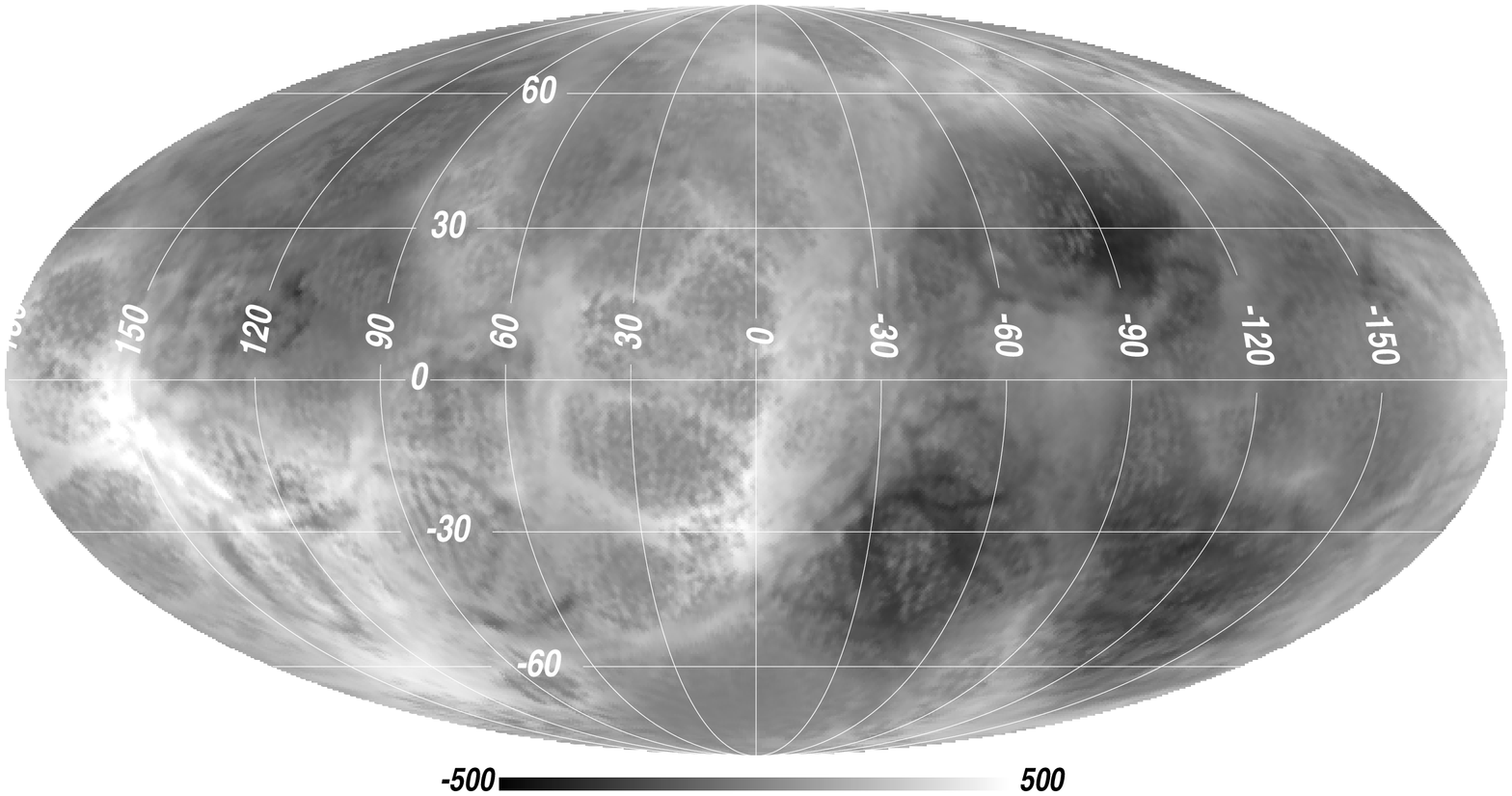}}
\vspace{-1.3cm}
\\
\rotatebox[]{90}{\includegraphics[width=3.5cm]{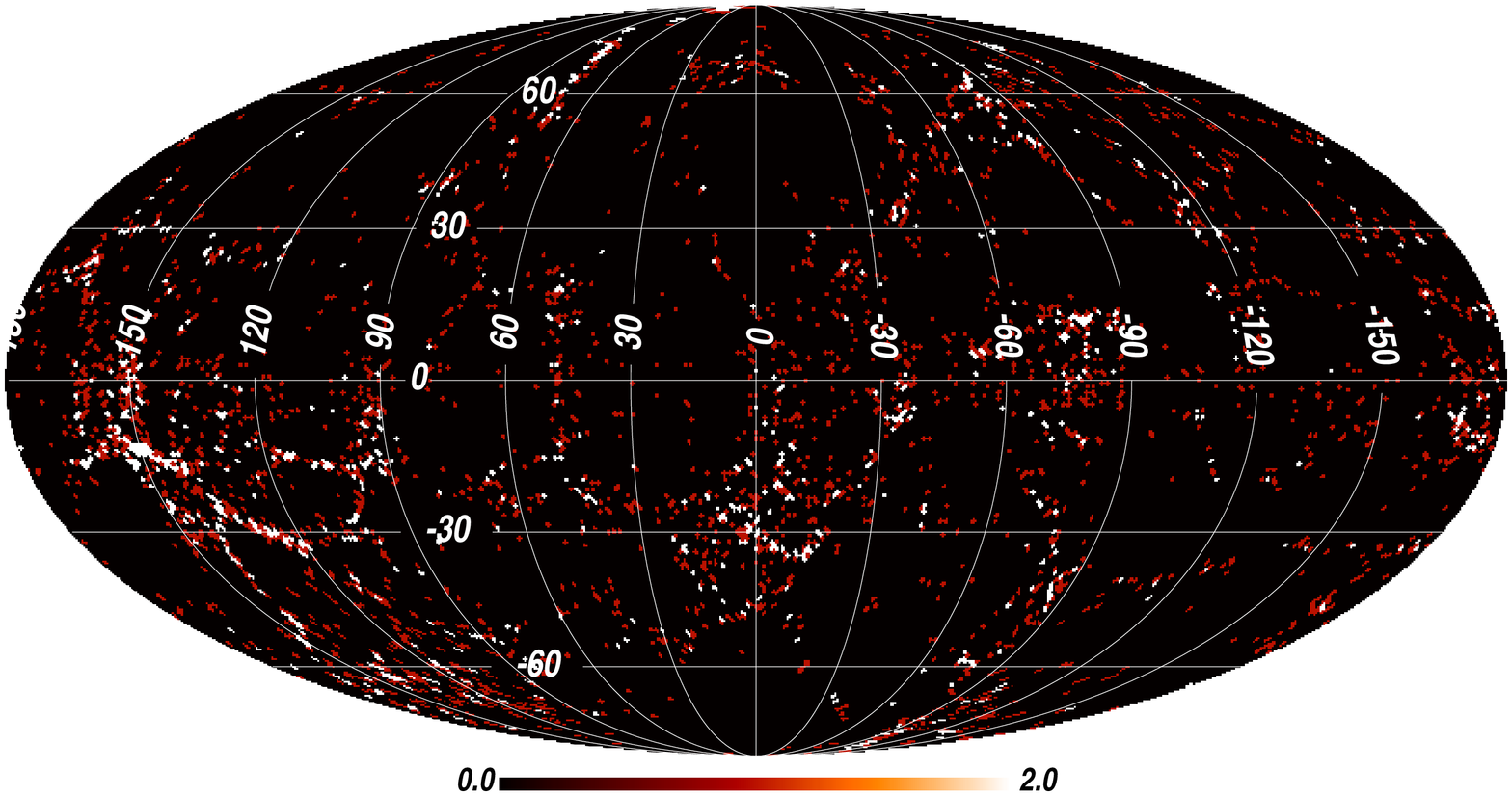}}
\put(60,121){\rm 50-60 $h^{-1}$ Mpc}
\rotatebox[]{90}{\includegraphics[width=3.5cm]{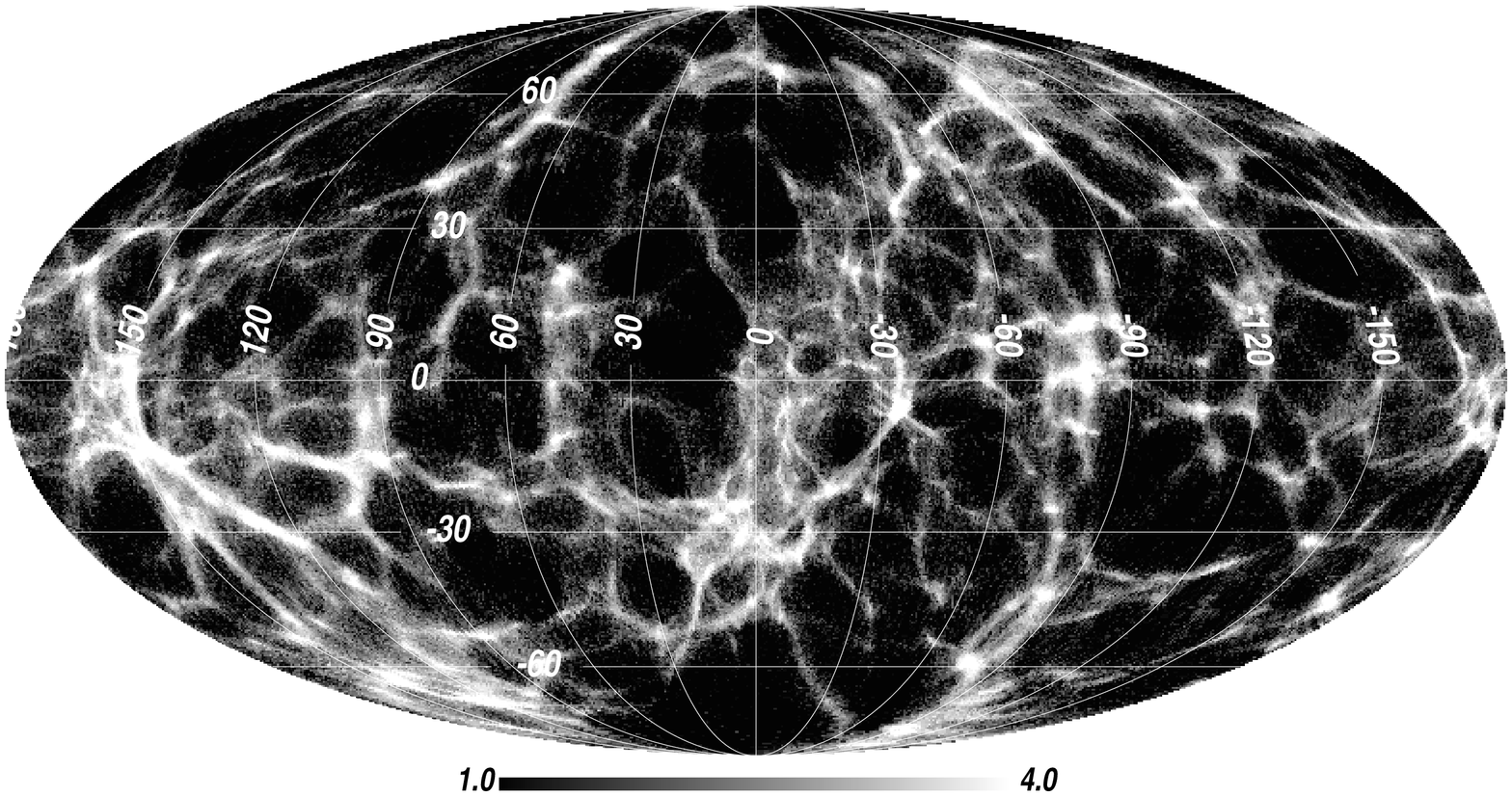}}
\rotatebox[]{90}{\includegraphics[width=3.5cm]{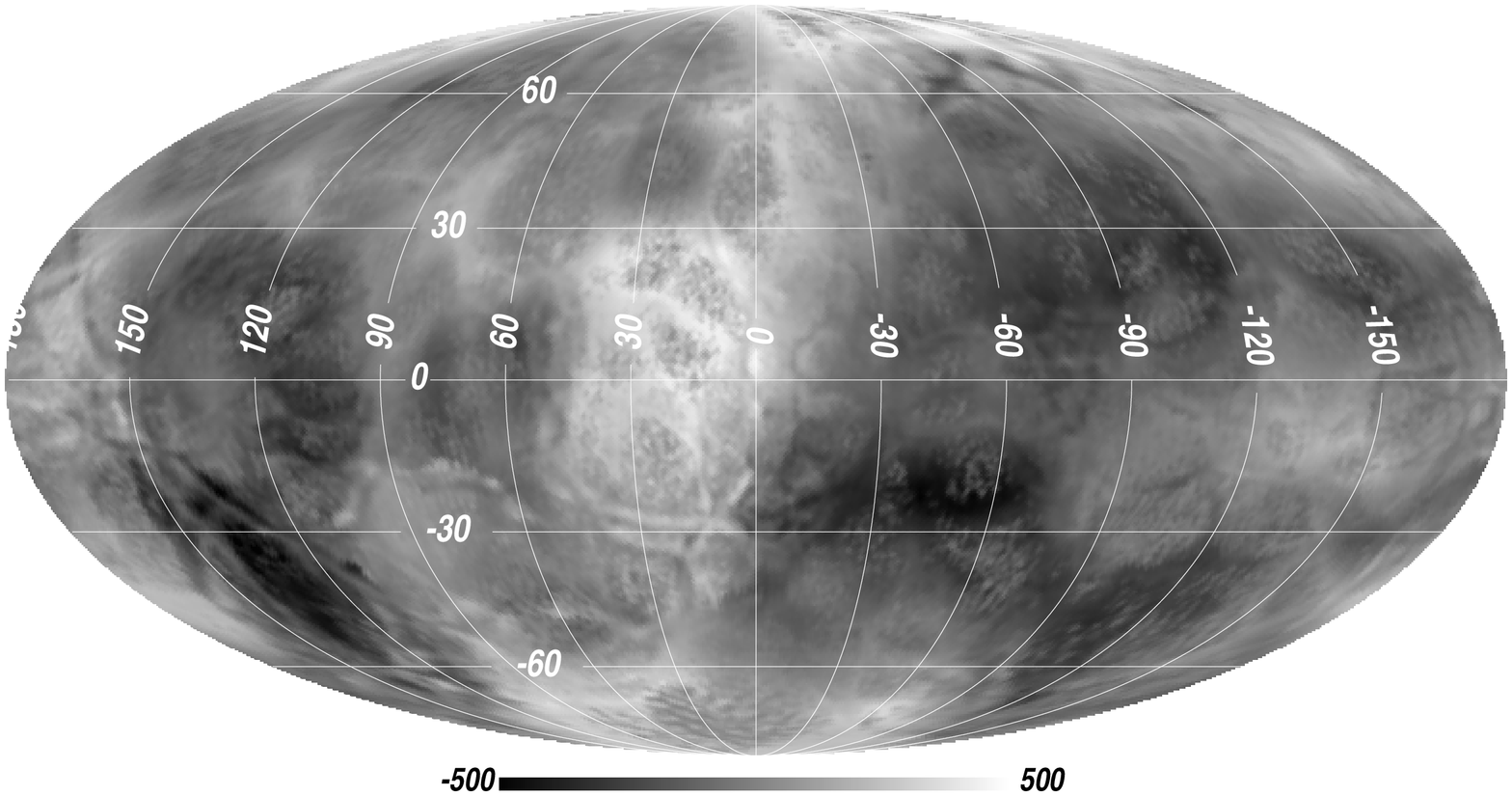}}
\vspace{-1.52cm}
\\
\rotatebox[]{90}{\includegraphics[width=3.5cm]{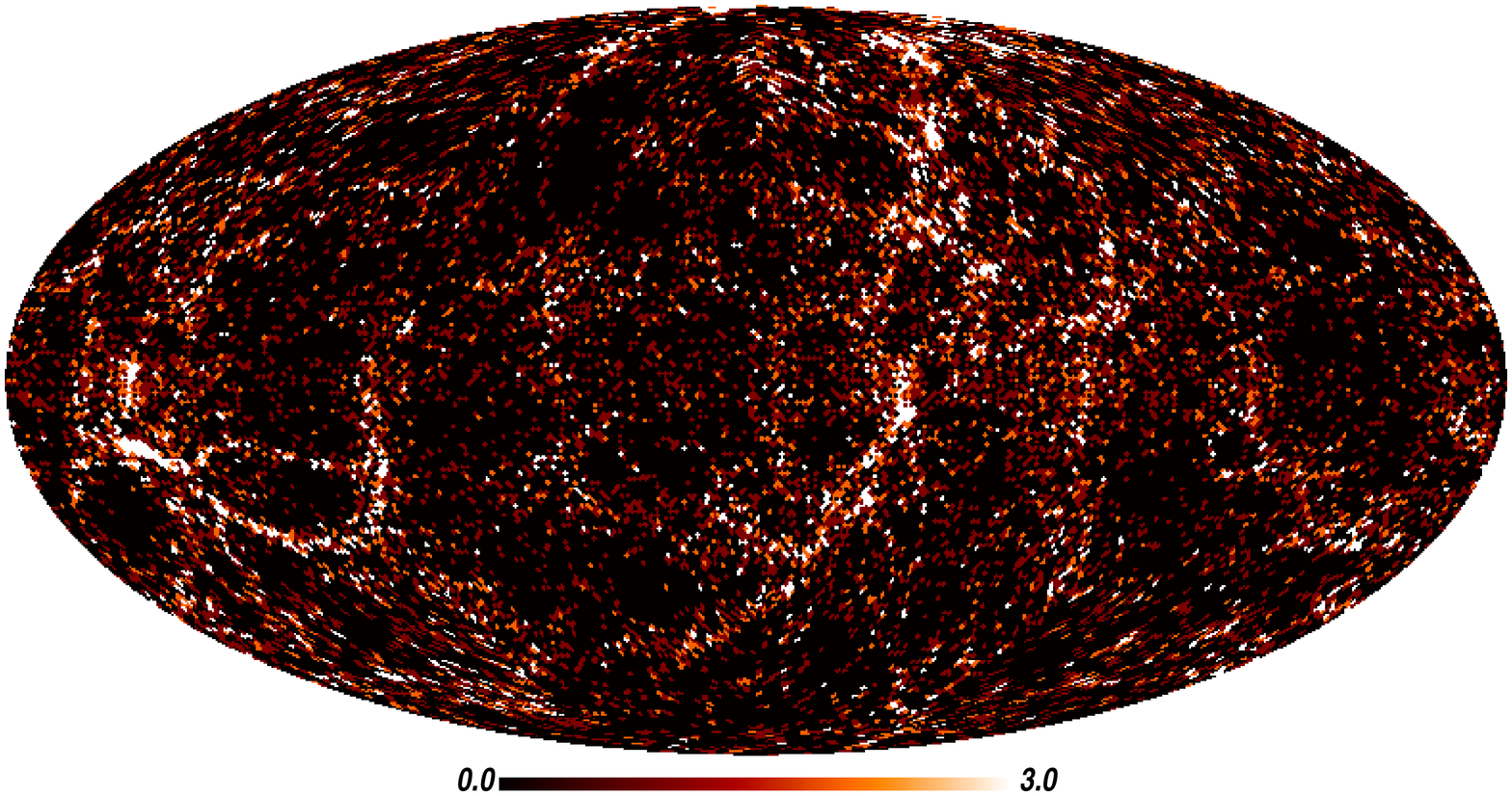}}
\hspace{0.4cm}
\rotatebox[]{90}{\includegraphics[width=3.5cm]{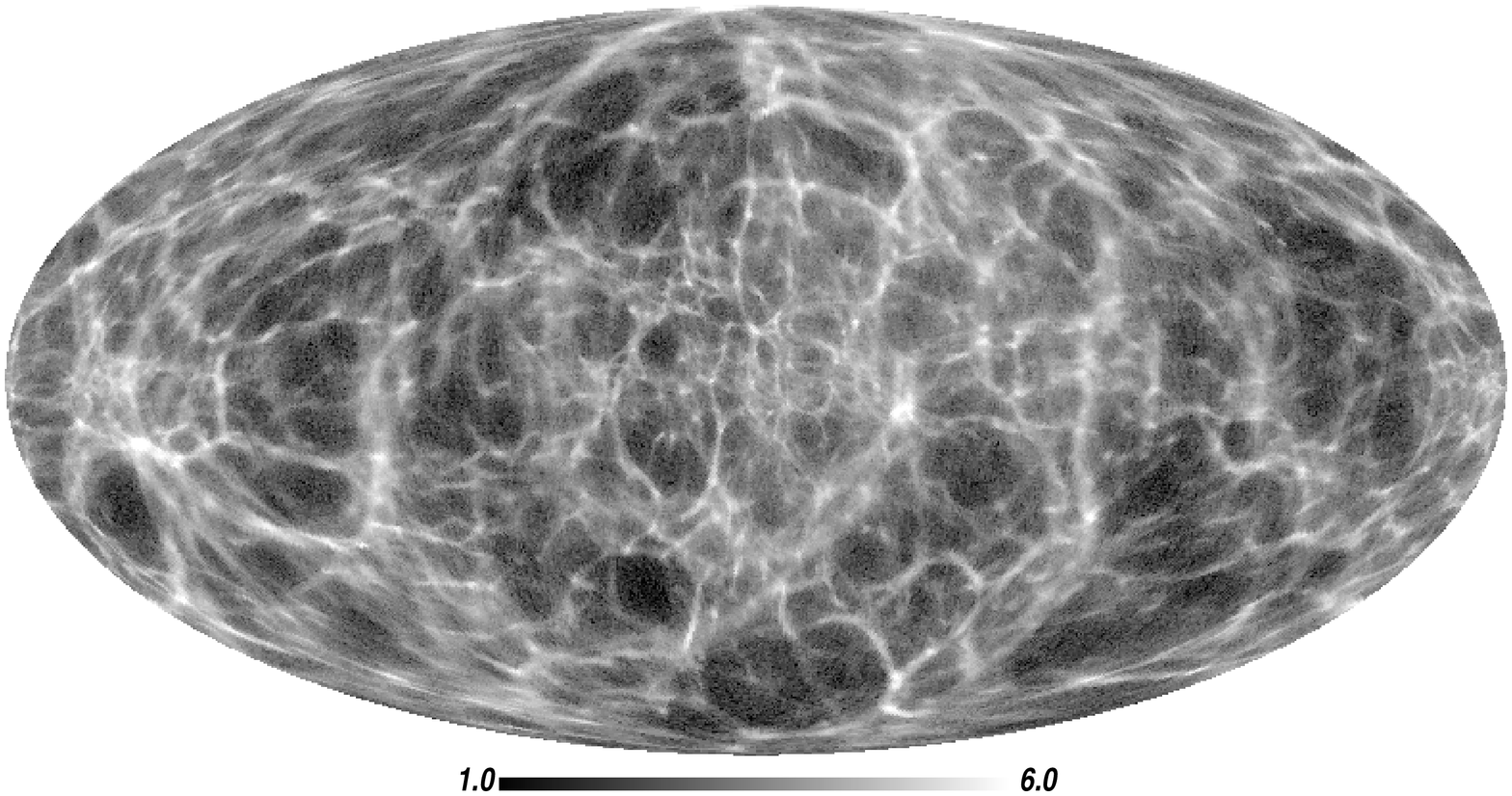}}
\put(-192,112.5){\rm 0-80 $h^{-1}$ Mpc}
\vspace{-1.48cm}
\end{tabular}
\caption{\label{fig:slices}  Mollview plots in Galactic coordinates using HEALPix  of the number counts of spectroscopic 2MASS galaxies. {\it  On the left} with $n_{\rm side}=64$ in different comoving distance ranges.  {\it The middle  panels} (right panel on the bottom) show the corresponding logarithm of number counts of particles from the reconstructed density field (one sample belonging to the highly correlated subsample) using $n_{\rm side}=128$. {\it The  right panels} (not present on the bottom) show the corresponding radial velocities averaged on each pixel with the number counts of particles in {\kms}  using $n_{\rm side}=128$. We note that the Galactic longitudes $l_{\rm gal:}$ have a shift of 360\degree.
} 
\end{figure*}

The analysis presented here is based on version 2.3 of the recently
released 2MASS redshift survey (2MRS), ${\rm K_s}=11.75$ catalogue 
\citep[][]{huchra}.
The 2MRS survey is unique in its sky coverage (91\%) and uniform
completeness (97.6\%) only limited 
near the Galactic plane ({\it the Zone of Avoidance}
(ZoA), where
$|b|<5^\circ$ and $|b|<8^\circ$ near the Galactic centre). 
 The ZoA could be sampled with a Poissonian likelihood describing the counts-in-cells of the galaxy distribution which would be limited to a coarse grid resolution \citep[see e.~g.][]{kitaura_log}.
To incorporate a mask treatment within a particle based reconstruction method like ours,  we would need to sample mock galaxies (or haloes) according to our structure formation model following schemes like the one proposed in \citet[][]{2002MNRAS.329..629S}, which is out of scope in this work.  
For the time being, we fill the ZoA
with random galaxies generated from the corresponding
longitude/distance bins in the adjacent
strips
\citep[][]{yahil}.
The method is robust for
the width of the 2MRS mask
 and has been thoroughly tested \citep[for details see][and references therein]{pirin_dipole}.
According to recent studies, the LG velocity should be independent of the treatment of the mask \citep[see][]{bilicki}.
Another essential ingredient is the radial selection function $f^{\rm sel}$ arising from the magnitude limit cut of the galaxy redshift survey. Here we derive $f^{\rm sel}$ from the 6 degree field galaxy
survey (6dFGS) luminosity function \citep[][]{jonesa}, imposing a
magnitude limit cut which corresponds to the one of 2MRS.
However, we note that the choice of the particular derivation of $f^{\rm sel}$ is not crucial \citep[see][]{2012arXiv1202.5206B}. 
The behaviour of $f^{\rm sel}$ can be further checked  in the power-spectra of the reconstructed initial fluctuations $P^{\rm rec}(k)$ (see below). 
 Finally, we compress the {\it fingers-of-god} after identifying them with a {\it friends-of-friends} algorithm taking into account the ellipsoidal distribution along the line-of-sight of groups of galaxies due to virial motions \citep[see][]{Tegmark-04}. The resulting catalog provides the Cartesian three-dimensional positions of the 2MRS galaxies  in Supergalactic coordinates including coherent redshift-space distortions.

\begin{figure}
\hspace{0.cm}
\includegraphics[width=8.5cm]{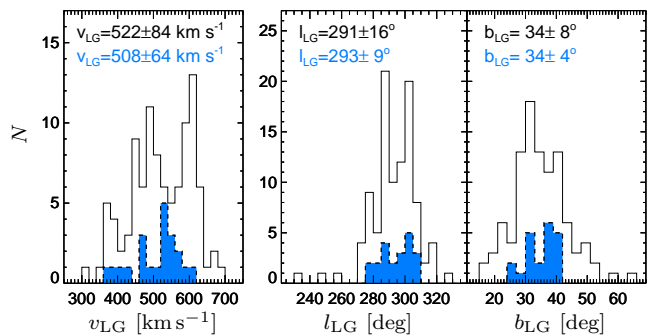}
\put(-247,73){\rotatebox[]{90}{\rm $N$}}
\put(-218,2){\rm $v_{\rm LG}$ [{\kms}]}
\put(-129,2){\rm $l_{\rm LG}$ [deg]}
\put(-57,2){\rm $b_{\rm LG}$ [deg]}
\vspace{-0.25cm}
\caption{\label{fig:stats}
 Histograms for the speed $v_{\rm LG}$ and the direction $(l_{\rm LG},b_{\rm LG})$ (Galactic) of the Local Group for the 100 reconstructed samples (black lines) and the highly correlated subsample  (see caption in Fig.~\ref{fig:slices})  (dashed lines filled in blue color code). The corresponding means and 1 sigma deviations are indicated.} 
\end{figure}

This catalog is used as an input for the \textsc{Kigen}-code which relies on 2LPT to describe structure formation
 \citep[see recent works on this subject:][and references therein]{kitaura_lin,kitaura_vel,jasche_2lpt,kitaura_kigen}.
We assume Gaussian initial density fields 
with a variance determined by
the cosmological parameters from the concordance $\Lambda$CDM--cosmology as provided by the Seven-Year Wilkinson Microwave Anisotropy Probe (WMAP) \citep[see table wmap7+bao+h0 in][]{wmap7}. 
 In this study we consider the data within a comoving box of 160 {\Muns} side (comprising about 30,000 galaxies).
Coherent redshift-space distortions are consistently modeled with 2LPT \citep[beyond the Kaiser limit;][]{Kaiser-87}  by adding a radial term  to the Eulerian real-space position $\mbi x=\mbi q+\mbi \Psi$, where $\mbi q$ is the Lagrangian position of the matter tracers at the initial conditions and $\mbi \Psi$ is the displacement field according to 2LPT. The redshift-space position $\mbi s$ of each matter tracer is thus given by the following equation in our model $\mbi s=\mbi q+\mbi \Psi+\mbi v_r$, with $\mbi v_r\equiv(\mbi v\cdot\hat{\mbi r})\hat{\mbi r}/(Ha)$, where $\mbi v$ is the full three dimensional 2LPT velocity field, $\hat{\mbi r}$ is the unit sight line vector, $H$ the Hubble constant and $a$ the scale factor. 
The \textsc{Kigen}-code  samples the initial fluctuations according to the set of Lagrangian test particles $\{\mbi q\}$ which under 2LPT yields a distribution in Eulerian redshift-space $\{\mbi s\}$ compatible with the observed galaxies $\{\mbi s_{\rm G}\}$. We use in each constrained 2LPT simulation $256^3$  test particles and compute the displacement field on a grid of $128^3$ cells with a resolution of $l_{\rm c}=1.25$ {\Muns}. Particles which are closer than $l_{\rm c}$ to a galaxy are considered to be ``friends'' of that galaxy and their Lagrangian positions are used as constraints to determine the initial Gaussian fluctuations  \citep[for more details on the method we refer to][]{kitaura_kigen}. 
We account for selection function effects and shot noise in our reconstruction by assigning in each iteration a weight to every particle according to a Poisson sample of the expected mass given by the inverse of $f^{\rm sel}$ evaluated at the distance of the corresponding galaxy.
 We consider for our analysis
 a set of 100 constrained 2LPT reconstructions of the LU, with the corresponding initial conditions and peculiar velocity fields. 
Our approach searches for Gaussian fields with a given prior power-spectrum (red curve in the right panel of Fig.~\ref{fig:corr}).  In Bayesian terms, the posterior from which we sample the initial density fluctuations is based on the above mentioned prior and a Poissonian likelihood for the test particles which we assume to be unbiased matter tracers. However, as the data are sparse and noisy and have a moderate galaxy bias \citep[see e.~g.][]{lavaux}, the posterior resulting from weighting the likelihood with the prior yields closely unbiased power spectra. This is also indicating that our treatment of the selection function is accurate. Otherwise a few modes on the very large scales would show a clear excess of power as we found in our tests. We note, that further studies should be done including a proper treatment of galaxy bias. 
{ It is difficult to do a fair comparison between the density field and the galaxy field due to bias and selection function effects. However,}  we find a remarkable correlation up to $k\sim1$ {\kuns} between the constrained 2LPT simulated overdensity field ($\delta^{\rm rec}_i\equiv N^{\rm part}_i/\overline{N}^{\rm part}-1$, with ${N}^{\rm part}$: test particles number count per cell $i$, $\overline{N}^{\rm part}$: mean) and the one directly computed from the galaxies ($\delta^{\rm gal}_i\equiv N^{\rm gal}_i/(\overline{N}^{\rm gal}f^{\rm sel}_i)-1$, with ${N}^{\rm gal}_i$: galaxy number count per cell $i$, $\overline{N}^{\rm gal}$: expected mean) (see left panel).  
The correlation does not vanish until $k\approx2.5$ {\kuns}. This result is supported by the Supergalactic plots in  Fig.~\ref{fig:SG}  (left and middle panels), which show how the nonlinear structures are accurately traced along the distribution of galaxies  at scales of 2-5 {\Muns}, even in regions with only of a few data points. The right panel shows the corresponding peculiar velocity field to a high level of detail demonstrating the formation of caustics in the high galaxy number density regions. { We expect to get better results using an improved structure formation model, especially in low density regions where 2LPT is known to perform worse \citep[see e.~g.][]{2002PhR...367....1B}. We leave such a work for future studies. We have checked that systematic discrepancies between the reconstructed 2LPT density field and the galaxy field are rather small (we obtain a statistical correlation coefficent of about 0.96 for the log-density field taking the volume of (160 {\Muns})$^3$) when one allows for an effective, linear bias combined with a Poisson sampling of the field that accounts for both galaxy bias and the failure of 2LPT on small scales. A way of improving this would consist on using the halo model as it was suggested by \citep[][]{2002MNRAS.329..629S}.}  
We use the ensemble of constrained simulations to compute the mean and standard deviation in each cell (see middle and right panel in Fig.~\ref{fig:CSG}). This gives us an estimate of the uncertainty in the position of the density peaks. If we look at Coma which central region has an extension of a few {\Muns} and is located at $(0,69,11)$ {\Muns} in SG-coordinates, we find that the uncertainty in the position is only of about 2-3  {\Muns}. The major uncertainties concern the extension around very massive structures being overall very robust (with S/N$>$1 in most of the cells, see right panel).
The quality of the reconstruction can be further assessed in the Mollview plots  of Fig.~\ref{fig:slices} performed with HEALPix \citep[][]{healpix}. Here the data (left panels) can be compared to the reconstruction (middle and right panels)  for different redshift slices. The accuracy of the reconstruction is very apparent to the level of connecting small number of galaxies through tiny filaments.  On the right panels the radial peculiar velocity field shows similar patterns than in linear theory \citep[see][for a careful description of the different structures]{erdogdu}, however reconstructing a significantly more complex structure. 
The bottom panels in Fig.~\ref{fig:slices} show the full projection on the sky to a distance of 80 {\Muns}. The data (left panel) have a very noisy appearance while the reconstruction (right panel) unveils  the corresponding cosmic web.\footnote{The figures are available in high resolution at: \url{www.aip.de/People/fkitaura/}}
 Fig.~\ref{fig:stats} shows our calculations of the LG velocity by evaluating the reconstructed velocity field at the center of the box. { The uncertainties in the estimation of the velocity field mainly depend on the randomness of the initial fluctuations and the shot noise due to the discrete galaxy sample including selection function effects. As it was shown in \cite{kitaura_vel} 2LPT is extremely accurate in the determination of the peculiar velocity field. We therefore do not expect significant systematic errors beyond the fact that we are considering the matter inside a volume of (160 {\Muns})$^3$.} We repeat our calculations for a subsample of extremely highly correlated reconstructions with the galaxy field finding consistent results and hence  demonstrating the robustness of our calculations.

\section{Discussion and conclusions}

We have performed a Bayesian cosmography analysis of the LU with the 2MRS galaxy redshift survey.
Our approach leads to more accurate estimates of the dynamics in the LU  with respect to previous ones  which either assume linear Eulerian relations  \citep[see e.~g.][]{fisher,erdogdu,bilicki} or  linear LPT  \citep[see e.~g.][]{lavaux_dipole}. 
 There are still a number of issues which should be further investigated, as for instance the impact of galaxy bias or full nonlinear evolution on the dynamics. { To extend the study presented here to larger volumes one should include a treatment of the so-called {\it rocket-effect} as it becomes more relevant in regions where the selection effects are more severe, i.~e. at larger distances to the observer \citep[see e.~g.][]{2012arXiv1202.5206B}. This effect could be iteratively corrected by re-computing the selection function at the galaxy estimated real-space position at each reconstruction step in our algorithm following the pioneering work of \citet[][]{yahil}.}
 Nevertheless, it is worth noting that  we model for the first time nonlocal and nonlinear effects in the calculation of the local cosmic flow including the tidal field component (within 2LPT). 
To perform higher order LPT it is essential to get accurate estimates of the linear component of the density field as it was pointed out in \citet[][]{kitaura_lin,kitaura_vel}. This is done in this work in a self-consistent way by iteratively sampling the Gaussian fields compatible with the data within 2LPT.  
In particular, we map local structures like the Local Void, the Local Supercluster (Virgo), the Coma Cluster, the Perseus Pisces Supercluster and the Great Attractor (Hydra and Centaurus) in great detail. 
 Our results show a wide range of LG velocity amplitudes (from about 350 to 600 {\kms}). These findings indicate  that the propagation of uncertainties is non-trivial and hence, the need for more studies including a nonlocal and nonlinear self-consistent treatment of gravity on different volumes \citep[as indicated in][]{pirin_dhalo}. 
The lower speeds than the ones from the CMB (16-18\% lower) found in our study and the angle separation of about 20\degree$\pm$10\degree indicate that the dipole has not converged yet considering the matter up to distances of about 80 {\Muns}. This is in good agreement with $\Lambda$CDM which predicts the dipole to have reached at those scales about 70 to 80\% of its total amplitude \citep[see e.~g.][]{lavaux_dipole,bilicki}.  Interestingly,  the amplitude and direction of the LG velocity we find is compatible within the (1 sigma) error bars with the direct observation of peculiar motions \citep[see][]{courtois}.  Our results are also consistent (within 2 sigma) with previous studies \citep[see e.~g.][]{pirin_dipole,lavaux_dipole,bilicki}.
We hope that the outcome of works like  the one presented here leads to a variety of applications ranging from cosmic web analysis, environmental studies and signal detections like the kinematic Sunyaev-Zel'dovich effect, to accurate constrained simulations enabling us to improve our understanding on structure formation in the LU.

{\bf Acknowledgements}
Thanks to S.~Hess { and E.~Branchini} for useful discussions. PE thanks the AIP for the hospitality. SEN is supported by the DFG grant MU1020 16-1, REA  by the ERC grant 246797 ``GALFORMOD'' and YH by the ISF (13/08).

{\small
\bibliographystyle{mn2e}
\bibliography{lit}
}

\end{document}